\documentclass{aastex}
\usepackage{emulateapj5}

\usepackage{natbib} 
\slugcomment{accepted for publication in ApJ Letters}

\shorttitle{Supergranulation Alignment in the Polar Regions}
\shortauthors{Nagashima et al.}

\begin{document}


\title{Detection of Supergranulation Alignment in Polar Regions of the Sun by Helioseismology}

\author{Kaori Nagashima\altaffilmark{1},
Junwei Zhao\altaffilmark{1}, Alexander G. Kosovichev\altaffilmark{1},
and Takashi Sekii\altaffilmark{2}}


\altaffiltext{1}{W.W. Hansen Experimental Physics Laboratory, 
452 Lomita Mall, Stanford University, Stanford, CA 94305-4085, USA; kaorin@stanford.edu}
\altaffiltext{2}{National Astronomical Observatory of Japan, 
2-21-1 Osawa, Mitaka, Tokyo 181-8588, Japan.}


\begin{abstract}
We report on a new phenomenon of `alignment' of 
supergranulation cells in the polar regions of the Sun.
Recent high-resolution datasets obtained by 
the Solar Optical Telescope onboard the Hinode satellite enabled
us to investigate supergranular structures in high-latitude regions 
of the Sun.
We have carried out a local helioseismology time-distance analysis of the data,
and detected acoustic travel-time variations due to the supergranular flows.
The supergranulation cells in both the north and south polar regions
show systematic alignment patterns in the north-south direction.
The south-pole datasets obtained in a month-long Hinode campaign
indicate that the supergranulation alignment property may be
quite common in the polar regions.
We also discuss the latitudinal dependence of the supergranulation
cell sizes; the data show that the east-west cell size decreases towards higher latitudes.
\end{abstract}



\section{Introduction}

Supergranules \citep{1954MNRAS.114...17H, 1962ApJ...135..474L}
are thought to be convection cells of the Sun
with the size of $20$--$30$ Mm and the lifetime of $\sim 1$ day.
Although more than half a century has passed since they were found,
their physical property is still puzzling and controversial.
\citet{1964ApJ...140.1120S} suggested that a convective instability associated with 
the He ionization zone might produce the supergranular cells.
\citet{2007AIPC..948..111S} argued that there is nothing special about the
supergranulation scale except that this is the scale favored by magnetic fields 
interacting with convective flows. 
\citet{2003Natur.421...43G}, \citet{2003ApJ...596L.259S}, and
\citet{2007ApJ...665L..75G} suggested that
supergranulation has properties of traveling convection waves,
while \citet{2004ApJ...608.1156R} and \citet{2004ApJ...608.1167L} discussed
that the supergranulation patterns may be organized by giant cells.
For recent discussions, see the review by 
\citet{2010LRSP....7....2R}.

Hydrodynamic processes, including supergranulation,
in the polar regions of the Sun most likely play an important role in the 
solar dynamics and magnetic activity cycles.
For example, \citet{2010GeoRL..3714107D} discussed that
the speed of the meridional flow in the polar regions could determine the
length of the activity cycle. 
The processes in the polar regions, however, have not been 
fully investigated from the observational point of view, mainly because 
observing the high-latitude regions from the ecliptic plane is difficult
due to severe foreshortening.
Numerical simulations of the turbulent convection on the Sun have suggested
some kind of latitudinal dependence of the supergranulation patterns,
although convection cells of the supergranulation scale
have not been clearly reproduced in the global simulations of the
convection zone, except in a shallow spherical shell study 
\citep{2002ApJ...581.1356D}.
The cells resolved in the global simulations of 
solar convection are mostly of a giant-cell scale. 
According to the recent simulation results
\citep[e.g.,][]{2002ApJ...570..865B, 2006ApJ...641..618M}, 
the convective cells in the lower latitude regions may align
parallel to the rotation axis.
In the higher-latitude regions, the structure of the cells looks more complicated
and less organized.

We take advantage of the Hinode high spatial resolution, and 
study the near-polar regions during the periods of the highest inclination of 
the solar axis to the ecliptic.
Our finding from the Hinode data analysis is that
the supergranulation cells in the high-latitude regions 
seem to align predominantly in the north-south direction.
This cannot be explained by the giant-cell scale structures
in the numerical simulations.

\section{Observations} \label{sec:PL_obs}


The Solar Optical Telescope (SOT; \citealt{2008SoPh..249..167T})
onboard the \textit{Hinode} satellite \citep{2007SoPh..243....3K}
has made several observing runs for local helioseismology. In particular,
the polar region of the southern hemisphere was observed on 2009 March 7, 
and the polar region in the northern hemisphere was observed 
on 2009 September 25. Since the inclination of the solar rotation axis
attains its maximum (about 7 degrees) in March and September every year, 
these are the best periods for observing the high-latitude regions.
For comparison and testing, we also obtained datasets of a region close to the east 
limb on 2009 September 27, 
and a region around the disk center on 2009 December 11.
All the observations were 16 hours long.

In March of 2010, SOT carried out nine helioseismology observing runs for the south polar region
approximately once every three days.
The observing period of each run was 12-16 hours, 
aside from the observation on March 16, which was 6.7 hours long.

The data acquired with a cadence of 60 seconds were Ca \textsc{ii} H line intensity maps as well as 
Fe \textsc{i} 557.6 nm intensity maps.
Since the field of view of the Fe-line maps obtained by the Narrowband Filter Imager
(NFI) of the SOT is wider than that of the Ca-line maps obtained by the Broadband Filter Imager (BFI), 
we set up the telescope pointing in such a way that 
the solar limb is in the NFI field of view.
The NFI data were then used for tracking the limb position and for determining 
the coordinates on the Sun in the BFI images.
For the helioseismology analysis
we used the Ca \textsc{ii} H data series. 

\section{Data Analysis} \label{sec:PL_dataanalysis}

We first aligned the images and determined the heliographic coordinates of
each image.
Then, we chose several points on the Sun within the SOT field of view,
as the central points for the Postel's projection (see below).
We tracked the central points with the local differential rotation rate \citep{1984SoPh...94...13S},
and remapped each image into the heliographic coordinates 
by using the Postel's projection.
Using the tracked and remapped images, we carried out the time-distance analysis
\citep{1997SoPh..170...63D, 1997ASSL..225..241K}
to measure variations of acoustic travel times and 
detect subsurface flow structures.

The SOT observations of the regions close to the limb 
were carried out with fixed telescope pointings.
Although the image stabilization system, 
the correlation tracker (CT; \citealt{2008SoPh..249..221S}), was turned on,
the field of view was gradually drifting toward the disk center
mainly due to the limb darkening. 
Therefore, we aligned the images by ensuring that
the limb stays in the same position in the field of view,
and then determined the solar coordinates of each image.

When we observe regions near the solar limb, 
foreshortening is significant, and appropriate remapping is required for
the data analysis.
We used the Postel's azimuthal equidistant projection \citep[e.g.,][]{1995ESASP.376b.147B,2008JPhCS.118a2090Z}. 
The projected map is what we would observe
by looking directly from above the region.
For the dataset taken at the disk center, however,
we did not project the images, because the distortion effect is negligible there.
Also, since for the disk center dataset
the SOT observed the region tracking its center
with the rotation rate of the Sun, we did not apply any further tracking. 

For the time--distance analysis,
cross-correlation functions of the surface wavefield of the 5-min oscillations were
calculated from the tracked and projected datasets.
We measured the acoustic (phase) travel times for several distances
by fitting a Gabor-type wavelet to the cross-correlation functions
\citep{1997ASSL..225..241K}.
For each distance $\Delta$, we averaged the oscillation signals over 
an annulus of the radius $\Delta$
around a target point, and then cross-correlated the averaged signals 
with the oscillation signal observed at the target point.
We obtained outward (from the target to the annulus) and 
inward (from the annulus to the target) travel times at each point 
in the field of view for a set of travel distances $\Delta$.
The difference of the outward and inward travel times indicates 
the diverging/converging flow around the target point
in the region of wave propagation below the surface. 
Note that for improving the signal-to-noise ratio we applied 
phase-speed filters following \citet{1997SoPh..170...63D}.

\section{Results} \label{sec:PL_results}

\subsection{Travel-time maps}

\begin{figure}[htpb]
\plotone{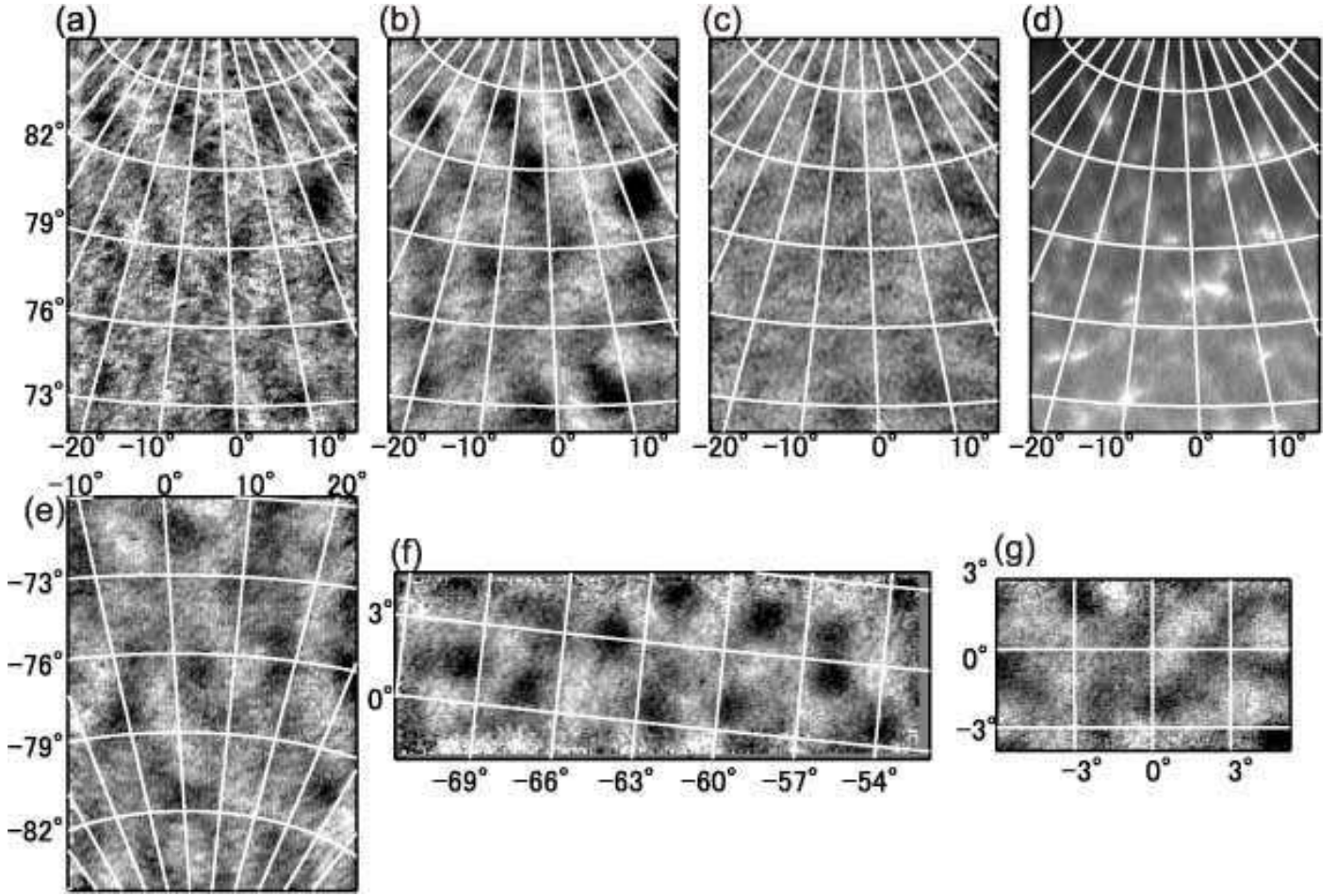}
\caption{
Outward--inward travel-time difference maps
of the north polar region on 2009 September 25,
for travel distances: (a) $8.4$ Mm, (b) $14.4$ Mm, and (c) $21.1$ Mm.  
Panel (d) shows a 4-hour averaged intensity image in the Ca \textsc{ii} H line
of the region. 
Panels (e), (f), and (g) show the outward--inward travel-time difference maps
of the south polar region on 2009 March 7, near the east limb on 2009 September 27, 
and around the disk center on 2009 December 11,
for the travel distances of $14.4$ Mm.
The gray scale used for the travel-time difference maps covers
the range from $-0.5$ minutes (black) to $+0.5$ minutes (white).
\label{fig:PL_tmapp4}}
\end{figure}

The outward--inward travel-time difference maps of the north polar region
are illustrated in Figure \ref{fig:PL_tmapp4}. 
Panels (a), (b), and (c) show the travel times 
for the distances, $\Delta$, of $8.4$, $14.4$, and $21.1$ Mm, 
(or $0.7^\circ$, $1.2^\circ$, and $1.7^\circ$) respectively. 
The black regions, where the outward travel times are shorter than the inward travel 
times, indicate diverging flows, while the white regions correspond to
converging flows.
The dark cells with white boundaries
are supergranular cells.
In the maps for the largest annuli (panel c),
the supergranular patterns are rather faint. 
Judging from the penetrating depths of the acoustic rays,
this indicates that the supergranules are not much deeper than, say,
$5$ Mm, which is consistent with other helioseismology results
\citep[e.g.,][]{ 2007ApJ...668.1189W,2007PASJ...59..637S}.
Hereafter, we use only the travel times for 
$\Delta=14.4$ Mm, because these travel-time maps 
(panel b) show the supergranular patterns most clearly.

Figure \ref{fig:PL_tmapp4}d shows a 4-hour averaged intensity image in the Ca \textsc{ii} H line
of the north polar region. 
It is well-known that supergranulation is seen 
as a bright network structure in the chromosphere
because magnetic field is concentrated at the supergranulation boundaries.
In this chromospheric image the network brightenings, though faint, are seen, 
and their distribution is consistent with the supergranular boundaries detected in 
the outward--inward travel-time maps (panels a to c).

\begin{figure}[htpb]
\epsscale{0.8}
\plotone{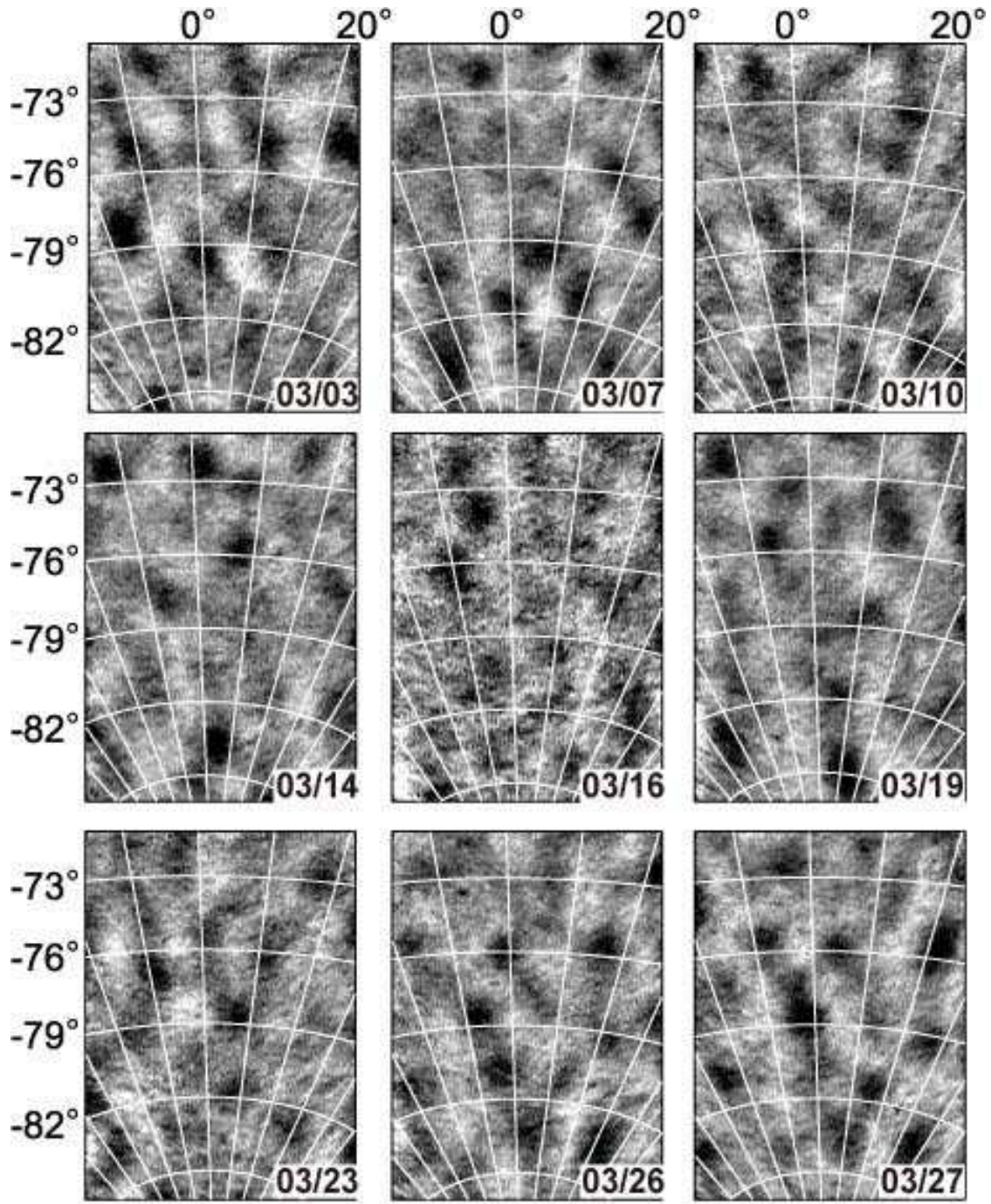}
\caption{Outward--inward travel-time difference maps of the south polar region
observed for 9 periods in March of 2010. The observing dates are indicated in the 
lower right corner. Note that the March 16 map is noisier than others because of 
short observation period. The gray scale is same as in Figure \ref{fig:PL_tmapp4}.\label{fig:PL_SP9}}
\end{figure}

In Figure \ref{fig:PL_tmapp4}, the outward--inward travel-time difference maps
of the south polar region (panel e), a region near the east limb (panel f), 
and a region around the disk center (panel g) are shown. 
Figure \ref{fig:PL_SP9} shows the nine outward--inward travel-time difference maps
of the south pole obtained for each of the nine observing runs in March of 2010. 
In these maps, the supergranular patterns in both polar regions 
and the east limb region are seen as clearly as in the travel time map of the disk center.
In the high-latitude region, however, the 
cells seem to align in the direction roughly from north to south, with some tilt.
The `alignment' patterns exist in both of the north and south polar regions;
Figure \ref{fig:PL_SP9} suggests that the `alignment' is
probably a characteristic structure in the polar regions.
There also seem to be areas with weak diverging flows between the aligned cells.
The alignment seems to be coherent for a distance of 2-3 cells. 


\subsection{Correlation of the travel-time difference maps}

\begin{figure}[htpb]
\epsscale{0.6}
\plotone{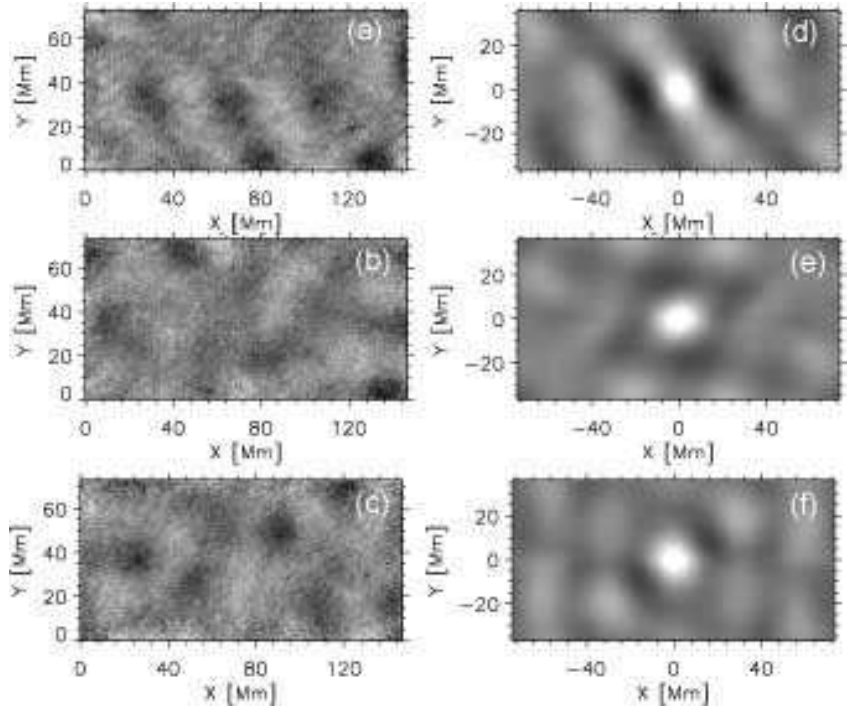}
\caption{Travel-time difference maps and two-dimensional correlation maps
of the travel-time difference.
Left column: Travel-time difference maps in (a) the north polar region,
(b) the disk center region, and (c) the east limb region. These maps 
are Postel's projected maps, and the sizes of the field-of-view are the same. 
Note that (a) and (c) is only a part of the full field of view.
The gray scale is same as in Figure \ref{fig:PL_tmapp4}.
Right column: Two-dimensional correlation maps of the travel-time difference maps
shown in the left panels. Panels (d), (e), and (f) are 
for the north polar region, the disk center region, and the east limb region,
respectively. The gray-scale range of the cross-correlation function is 
from $-0.5$ (black) to $+0.5$ (white).
\label{fig:PL_cor}}
\end{figure}

For a quantitative characterization of the alignment,
we calculated two-dimensional 
correlations of the travel-time difference maps. 
Figure \ref{fig:PL_cor} shows the correlation functions of the travel-time difference maps
for the north polar region and the lower-latitude regions as well. 
The correlation function of the north polar datasets (panel d) 
does indicate a striking alignment in the map, i.e., the correlation is the strongest in the
northeast-southwest direction. 
On the other hand, for the disk center and the east limb region maps
(panels e and f)
the correlation shows no particular preference in direction; namely, 
it shows a nearly random distribution.

\begin{figure}[htpb]
\epsscale{0.3}
\plotone{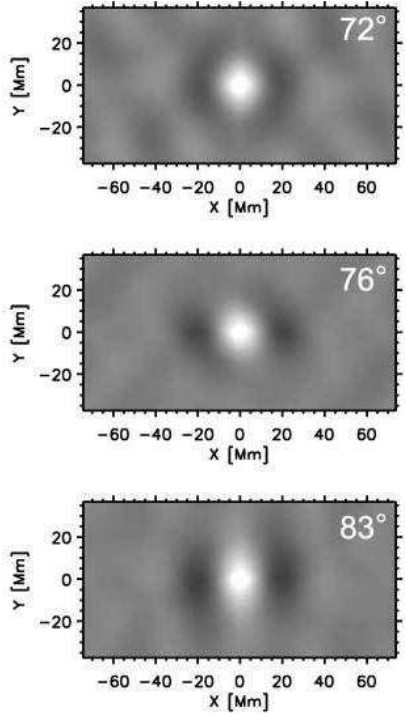}
\caption{Averages of the two-dimensional cross-correlation functions over all 
the nine south pole datasets for 
the central latitudes: $72^{\circ}$ (top), $76^{\circ}$ (middle), and $83^{\circ}$ (bottom).
The gray scale is same as in the right panels of Figure \ref{fig:PL_cor}. \label{fig:PL_corav}}
\end{figure}

Actually, the strength of the `alignment' differs from dataset to dataset.
The alignment, however, does exist in the averaged correlation functions
of the travel-time difference maps, calculated for different latitudes
(Figure \ref{fig:PL_corav}). The alignment is most notable at
the highest latitude, as shown in the bottom panel.
The alignment is almost in the north-south direction, but not precisely.
We estimated the mean direction and its variance for the nine datasets.
We measured the peak position angle along a 20-Mm circle centered at the origin of
the correlation maps,
and found that the mean and the variance alignment angles are
$90.1 \pm 37.0$ degrees at $72^{\circ}$ in latitude,
$102.5 \pm 23.2$ degrees at $76^{\circ}$, and
$92.1 \pm 11.5$ degrees at $83^{\circ}$.
Note that here the direction of $0$ degree is to the west,
and $90$ degrees is to the north.
This analysis indicates: 1) the alignment is, on average, in the nearly north-south direction,
and 2) the variance of the alignment angle is smaller in the 
higher latitude regions.
On a short time scale, the alignment might be affected by the polar dynamics,
such as varying differential rotation \citep{1998SoPh..179....1Y}, 
meridional flow, and a polar vortex \citep[e.g.,][]{1979ApJ...231..284G}.

The correlation functions of the travel-time maps also show us variation of the 
cell size with latitude.
We have found that the east-west size of the supergranular cells 
in the high-latitude regions tends to be smaller than in the 
low-latitude regions. We define the cell radius (a half of the full size) as 
a correlation length: the distance in $x$ (east-west) direction
from the origin to the point where the correlation is down to zero 
in the correlation map. 
The cell correlation radii are 10-12 Mm in the north/south polar regions,
and 13-14 Mm in the east limb region and around the disk center. 
Moreover, from the average of the correlation function over the nine south polar
datasets, we obtained the cell radii as 
$12.0 \pm 0.4$ Mm around $72^{\circ}$ in latitude,
$11.8 \pm 0.2$ Mm around $76^{\circ}$, and $11.2 \pm 0.2$ Mm around $83^{\circ}$. 
The errors are estimated from the variance of the nine datasets.
This implies that the cell size depends on latitude.
This type of latitudinal dependence is consistent with the results 
obtained from f-mode helioseismology analysis by \citet{2008SoPh..251..417H} 
and from Ca \textsc{ii} K line observations of the chromospheric network
\citep{1982SoPh...75...75B, 1989A&A...213..431M}, although 
these observations were for mid- or lower-latitude regions.
Note that by this method we could not define the cell size in the north-south direction 
because the correlation function does not drop to zero
in this direction (see Figure \ref{fig:PL_corav}).
This might be related to the fact that regions with weak diverging flows 
are located between the aligned cells in the north-south direction
as can be seen in the travel-time maps (Figures \ref{fig:PL_tmapp4} and \ref{fig:PL_SP9}),
but this suggestion needs further investigation.

We verified that the alignment is not due to some kind of artifacts
caused by the limb observation, because the alignment is not found
in the control study of the east limb region. In general,
any kind of limb-observation effect would not affect positions of the cells 
in the travel-time maps, thereby producing a spurious alignment.
Also, it is unlikely that projection effects affected our
estimates of the cell size in the east-west direction.


\subsection{Temporal evolution}

\begin{figure}[htpb]
\epsscale{0.6}
\plotone{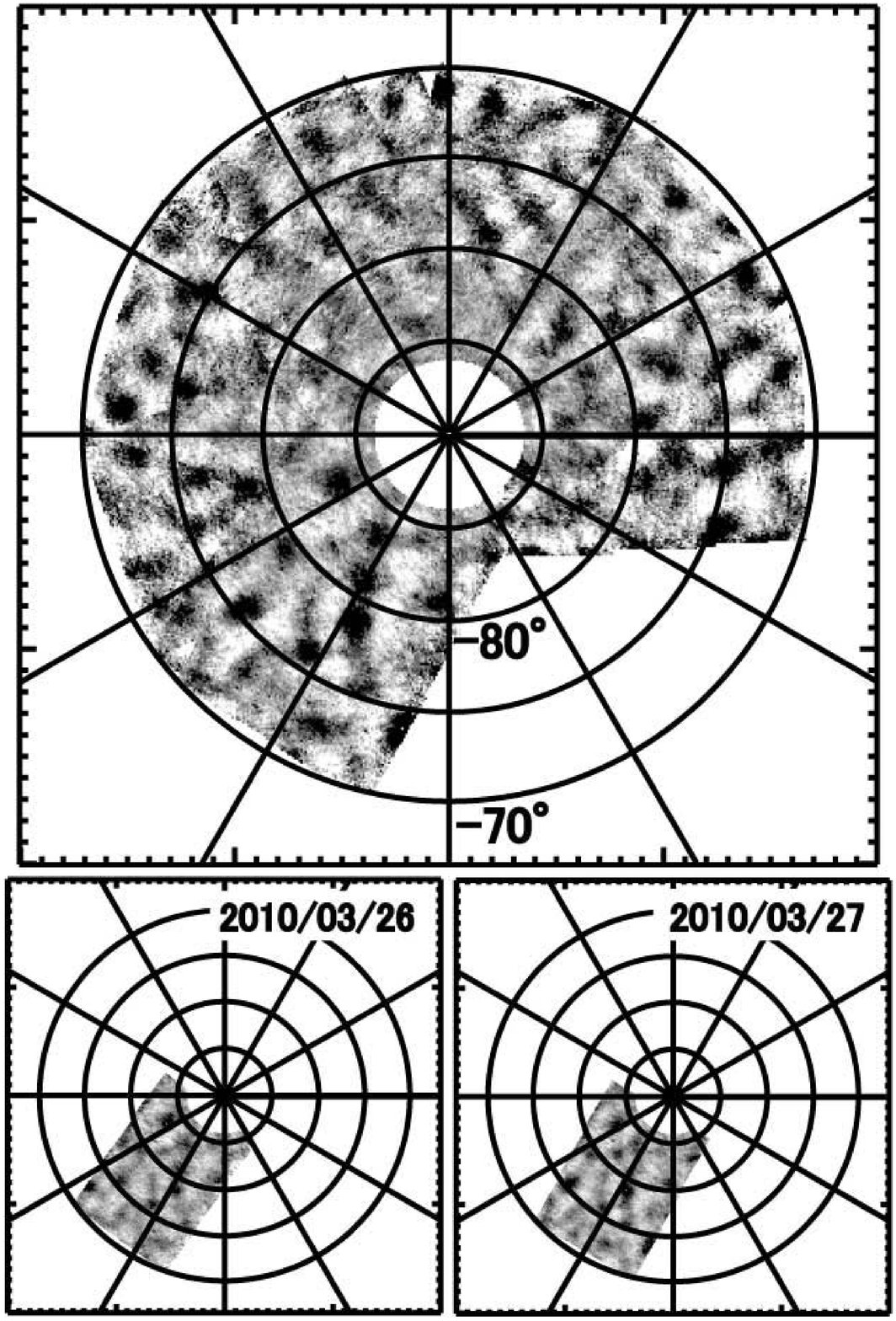}
\caption{Outward--inward travel-time difference maps of 
the south polar region observed in March of 2010. 
The upper panel shows the combined polar view 
above the South Pole, obtained by overlapping the results of the 
nine observing runs.
The lower panels are the last two 
consecutive travel-time difference maps.
The concentric circles are the latitudinal lines spaced by five degrees,
while the straight lines are the longitudinal lines spaced by 30 degrees.
The gray scales used for the upper panel and the lower panels
cover the range from $-0.25$ minutes to $+0.25$ minutes
and the range from $-0.5$ minutes to $+0.5$ minutes, respectively.
\label{fig:PL_SPall}}
\end{figure}

The upper panel of Figure \ref{fig:PL_SPall} 
shows the combined outward--inward travel-time difference map 
of the nine south polar datasets of March of 2010. 
This is a view from above the South Pole of the Sun.
The first observation of March 3 is the rightmost one,
and the following observations are located sequentially
counter-clockwise. We calculated the longitudinal difference 
between the observing dates by using the differential rotation rate \citep{1984SoPh...94...13S}
at the central latitude of each observation.
Note that the supergranulation pattern in Figure \ref{fig:PL_SPall} looks 
fainter than in the individual maps (Figure \ref{fig:PL_SP9})
because of overlapping of the different time intervals.

The lower panels show the last two datasets,
the temporal separation between which was only 15.5 hours,
and their fields of view significantly overlapped. 
Many features in the overlapping region look similar;
a clear example is along the line at angle of 240 degrees.
Obviously, the cells and their alignment pattern survive 
at least 15.5 hours, which is consistent with the known fact that 
the supergranular cell lifetime is about a day. 
For a detailed study of the temporal evolution
we need more consecutive datasets. 

\section{Discussions} \label{sec:PL_disc}

We have found evidence indicating that:
1) supergranular cells tend to align roughly in the north-south direction
(with some tilt) in the polar regions; the alignment is coherent over 
the scale of a few cells, 
2) there may be areas with weak diverging flows between the aligned cells, and
3) the horizontal (east-west) size of the cells decreases towards higher latitudes.

\citet{2004ApJ...608.1167L} found observational evidence of the alignment 
of supergranular cells in low latitude regions below 60 degrees, and 
suggested that the alignment is controlled by giant cells, because converging
flows at the giant cell boundaries are expected to align parallel to the rotation axis 
at low latitudes, as predicted by numerical simulations 
\citep{2002ApJ...570..865B, 2006ApJ...641..618M}.
The alignment reported by \citet{2004ApJ...608.1167L} was found
during 8-day observations, while our observing runs were only 16 hours long.
Since the lifetime of a supergranular cell is 
about a day,
the alignment found by them is somewhat statistical, i.e.,
it does not mean that supergranules are aligned at a given time,
as was found by our study.
Yet, in both cases the alignment may be a result of giant-cell flows
organizing smaller-scale flow fields.
What we essentially need for further study is
to determine the lifetime of the organized structures, and to
compare with the expected giant-cell lifetime (about a month),
as well as to study the relationship with the global magnetic structure of the Sun.
Also, it should be interesting to compare the polar supergranulation pattern with
the global supergranulation pattern by using the latest helioseismology instrument, 
Helioseismic and Magnetic Imager (HMI) onboard the Solar Dynamics Observatory.

Perhaps, the fact that the cells in the polar regions seem to have areas with 
weak diverging flows between the aligned cells can be explained by  
the convective rolls. If there are elongated cells in the north-south direction
with counter-rotating convective motions, the outward--inward travel-time 
difference maps would show these patterns. 
However, the theory \citep[e.g.][]{1970JFM....44..441B} predicts the 
convective roll patterns in low-latitude regions
of the convection zone below the surface, which remain to be detected.



\acknowledgments

Hinode is a Japanese mission developed and launched by ISAS/JAXA,
with NAOJ as domestic partner and NASA and STFC (UK) as international partners.
It is operated by these agencies in co-operation with ESA and NSC (Norway).
This work was partly carried out at the NAOJ Hinode Science Center,
which is supported by the Grant-in-Aid for 
Creative Scientific Research 
``The Basic Study of Space Weather Prediction'' from MEXT, Japan 
(Head Investigator: K.~Shibata), generous donations 
from Sun Microsystems, and NAOJ internal funding.
Part of this work was done while 
K.~Nagashima had been supported by the Research Fellowship from the 
Japan Society for the Promotion of Science for Young Scientists.
This research was supported by NASA grants NNX09AB10G and NNX09AG81G.



{\it Facility:} \facility{Hinode (SOT)}

\begin{thebibliography}{30}

\bibitem[{{Bogart} {et~al.}(1995){Bogart}, {S\'a}, {Duvall}, {Haber}, {Toomre}, \&
  {Hill}}]{1995ESASP.376b.147B}
{Bogart}, R.~S., {S\'a}, L.~A.~D., {Duvall}, T. L., {Haber}, D.~A., {Toomre}, J., \& {Hill}, F.
  1995, in ESA Special Publication, Vol. 376, Helioseismology, 147--150

\bibitem[{{Brun} \& {Toomre}(2002)}]{2002ApJ...570..865B}
{Brun}, A.~S., \& {Toomre}, J. 2002, \apj, 570, 865

\bibitem[{{Brune} \& {W{\"o}hl}(1982)}]{1982SoPh...75...75B}
{Brune}, R., \& {W{\"o}hl}, H. 1982, \solphys, 75, 75

\bibitem[{{Busse}(1970)}]{1970JFM....44..441B}
{Busse}, F.~H. 1970, Journal of Fluid Mechanics, 44, 441

\bibitem[{{De Rosa} {et~al.}(2002){De Rosa}, {Gilman}, \&
  {Toomre}}]{2002ApJ...581.1356D}
{De Rosa}, M.~L., {Gilman}, P.~A., \& {Toomre}, J. 2002, \apj, 581, 1356

\bibitem[Dikpati {et~al.}(2010)]{2010GeoRL..3714107D} Dikpati, M., Gilman, 
P.~A., de Toma, G., \& Ulrich, R.~K.\ 2010, \grl, 37, 14107 

\bibitem[{{Duvall} {et~al.}(1997)}]{1997SoPh..170...63D}
{Duvall}, Jr., T.~L., {et~al.} 1997, \solphys, 170, 63

\bibitem[Gilman(1979)]{1979ApJ...231..284G} Gilman, P.~A.\ 1979, 
\apj, 231, 284

\bibitem[{{Gizon} {et~al.}(2003){Gizon}, {Duvall}, \&
  {Schou}}]{2003Natur.421...43G}
{Gizon}, L., {Duvall}, T.~L., \& {Schou}, J. 2003, \nat, 421, 43

\bibitem[{{Green} \& {Kosovichev}(2007)}]{2007ApJ...665L..75G}
{Green}, C.~A., \& {Kosovichev}, A.~G. 2007, \apjl, 665, L75

\bibitem[{{Hart}(1954)}]{1954MNRAS.114...17H}
{Hart}, A.~B. 1954, \mnras, 114, 17

\bibitem[{{Hirzberger} {et~al.}(2008){Hirzberger}, {Gizon}, {Solanki}, \&
  {Duvall}}]{2008SoPh..251..417H}
{Hirzberger}, J., {Gizon}, L., {Solanki}, S.~K., \& {Duvall}, T.~L. 2008,
  \solphys, 251, 417

\bibitem[{{Kosovichev} \& {Duvall}(1997)}]{1997ASSL..225..241K}
{Kosovichev}, A.~G., \& {Duvall}, Jr., T.~L. 1997, in Astrophysics and Space
  Science Library, Vol. 225, SCORe'96 : Solar Convection and Oscillations and
  their Relationship, ed. F.~P. {Pijpers}, J.~{Christensen-Dalsgaard}, \& C.~S.
  {Rosenthal} (Dordrecht: Kluwer), 241

\bibitem[{{Kosugi} {et~al.}(2007)}]{2007SoPh..243....3K}
{Kosugi}, T., {et~al.} 2007, \solphys, 243, 3

\bibitem[{{Leighton} {et~al.}(1962){Leighton}, {Noyes}, \&
  {Simon}}]{1962ApJ...135..474L}
{Leighton}, R.~B., {Noyes}, R.~W., \& {Simon}, G.~W. 1962, \apj, 135, 474

\bibitem[{{Lisle} {et~al.}(2004){Lisle}, {Rast}, \&
  {Toomre}}]{2004ApJ...608.1167L}
{Lisle}, J.~P., {Rast}, M.~P., \& {Toomre}, J. 2004, \apj, 608, 1167

\bibitem[{{Miesch} {et~al.}(2006){Miesch}, {Brun}, \&
  {Toomre}}]{2006ApJ...641..618M}
{Miesch}, M.~S., {Brun}, A.~S., \& {Toomre}, J. 2006, \apj, 641, 618

\bibitem[{{M{\"u}nzer} {et~al.}(1989){M{\"u}nzer}, {Hanslmeier}}, {Schr{\"o}ter},
  \& {W{\"o}hl}]{1989A&A...213..431M}
{M{\"u}nzer}, H., {Hanslmeier}, A., {Schr{\"o}ter}, E.~H., \& {W{\"o}hl}, H.  
  1989, \aap, 213, 431

\bibitem[{{Rast} {et~al.}(2004){Rast}, {Lisle}, \&
  {Toomre}}]{2004ApJ...608.1156R}
{Rast}, M.~P., {Lisle}, J.~P., \& {Toomre}, J. 2004, \apj, 608, 1156

\bibitem[{{Rieutord} \& {Rincon}(2010)}]{2010LRSP....7....2R}
{Rieutord}, M., \& {Rincon}, F. 2010, Living Reviews in Solar Physics, 7, 2

\bibitem[{{Schou}(2003)}]{2003ApJ...596L.259S}
{Schou}, J. 2003, \apjl, 596, L259

\bibitem[{{Sekii} {et~al.}(2007)}]{2007PASJ...59..637S}
{Sekii}, T., {et~al.} 2007, \pasj, 59, S637

\bibitem[{{Shimizu} {et~al.}(2008)}]{2008SoPh..249..221S}
{Shimizu}, T., {et~al.} 2008, \solphys, 249, 221

\bibitem[{{Simon} \& {Leighton}(1964)}]{1964ApJ...140.1120S}
{Simon}, G.~W., \& {Leighton}, R.~B. 1964, \apj, 140, 1120

\bibitem[{{Snodgrass}(1984)}]{1984SoPh...94...13S}
{Snodgrass}, H.~B. 1984, \solphys, 94, 13

\bibitem[{{Stein} {et~al.}(2007){Stein}, {Benson}, {Georgobiani}, {Nordlund},
  \& {Schaffenberger}}]{2007AIPC..948..111S}
{Stein}, R.~F., {Benson}, D., {Georgobiani}, D., {Nordlund}, {\AA}., \&
  {Schaffenberger}, W. 2007, in American Institute of Physics Conference
  Series, Vol. 948, Unsolved Problems in Stellar Physics: A Conference in Honor
  of Douglas Gough, ed. {R.~J.~Stancliffe, G.~Houdek, R.~G.~Martin, \&
  C.~A.~Tout}, 111--115

\bibitem[{{Tsuneta} {et~al.}(2008)}]{2008SoPh..249..167T}
{Tsuneta}, S., {et~al.} 2008, \solphys, 249, 167

\bibitem[{{Woodard}(2007)}]{2007ApJ...668.1189W}
{Woodard}, M.~F. 2007, \apj, 668, 1189

\bibitem[Ye \& Livingston(1998)]{1998SoPh..179....1Y} 
Ye, B., \& Livingston, W.\ 1998, \solphys, 179, 1 

\bibitem[{{Zaatri} {et~al.}(2008){Zaatri}, {Corbard}, {Roth}, {Gonz{\'a}lez
  Hern{\'a}ndez}, \& {von der L{\"u}he}}]{2008JPhCS.118a2090Z}
{Zaatri}, A., {Corbard}, T., {Roth}, M., {Gonz{\'a}lez Hern{\'a}ndez}, I., \&
  {von der L{\"u}he}, O. 2008, Journal of Physics Conference Series, 118,
  012090

\end{thebibliography}
\end{document}